# Superposition of intra- and inter-layer excitons in twistronic MoSe$_2$/WSe$_2$ bilayers probed by resonant Raman scattering

*Liam P. McDonnell[1], Jacob J.S. Viner[1], David A. Ruiz-Tijerina[2], Pasqual Rivera[3], Xiaodong Xu[3], Vladimir I. Fal'ko[4,5] and David C. Smith[1*]*

[1] School of Physics and Astronomy, University of Southampton, Southampton SO17 1BJ, United Kingdom.

[2] Centro de Nanociencias y Nanotecnología, Universidad Nacional Autónoma de México, C.P. 22800, Ensenada, Baja California, México.

[3] Department of Physics, University of Washington, Seattle, WA, USA.

[4] National Graphene Institute, University of Manchester, M13 9PL, United Kingdom.

[5] Henry Royce Institute for Advanced Materials, University of Manchester, Manchester, M13 9PL, United Kingdom.

*Corresponding author – D.C.Smith@soton.ac.uk

**Hybridisation of electronic bands of two-dimensional materials, assembled into twistronic heterostructures, enables one to tune their optoelectronic properties by selecting conditions for resonant interlayer hybridisation. Resonant interlayer hybridisation qualitatively modifies the excitons in such heterostructures, transforming these optically active modes into superposition states of interlayer and intralayer excitons. For MoSe$_2$/WSe$_2$ heterostructures, strong hybridization occurs between the holes in the spin-split valence band of WSe$_2$ and in the top valence band of MoSe$_2$, especially when both are bound to the same electron in the lowest conduction band of WSe$_2$. Here we use resonance Raman scattering to provide direct evidence for the hybridisation of excitons in twistronic MoSe$_2$/WSe$_2$ structures, by observing scattering of specific excitons by phonons in both WSe$_2$ and MoSe$_2$. We also demonstrate that resonance Raman scattering spectroscopy opens up a wide range of possibilities for quantifying the layer composition of the superposition states of the exciton and the interlayer hybridisation parameters in heterostructures of two-dimensional materials.**

Monolayers of transition-metal dichalcogenides (TMD) are semiconductors with a plethora of appealing optical properties [1–3]. Unlike the bulk compounds, they feature a direct band gap and a strong light-matter interaction involving optical transitions between the K-valleys electrons and holes [4,5]. Angular momentum, implicit in the conduction and valence band K-valleys states that are dominated by transition metal d-orbitals, determines a valley-selective coupling to the circularly polarised light. At the same time, spin-orbit coupling in transition-metals sets a tight spin-valley locking for the valence band holes, leading to the spin-selectivity of the optical transitions in each valley[2].

In van der Waals heterostructures of TMDs [8], these properties are additionally enriched by the layer-indirect optical transitions[9] across the type-II interface. These excitations are much more sensitive to twist angle than the intralayer excitons. Changing between parallel ($\theta \approx 0^o$) and anti-parallel layer ($\theta \approx 60^o$) alignment changes the energy of bright interlayer excitons by the spin-orbit splitting of carrier in the 'rotated' layer (see Fig 1a). Also, for twist angles near (anti)parallel alignment, a twist-dependent momentum mismatch between K-valleys means that optically active interlayer excitons (IX) have a finite centre of mass momentum and substantial kinetic energy creating the conditions for valley memory[9], tunability of the lifetime, accumulation of high densities of IX, and even Bose-Einstein condensation[10] at low temperatures.

Interlayer tunnelling of electrons and holes also leads to the hybridisation of the intralayer and interlayer excitons. This can lead to strong hybridisation when energies of the two states are resonant [8]. Such resonant conditions naturally occur in homobilayers and few-layer films of TMDs [9,10]. For some heterostructures, such as highly aligned $MoSe_2/WS_2$ bilayers, strong hybridisation has been observed[11] between the nearly degenerate lowest-energy IX and the $MoS_2$ A exciton. Evidence for the resonance anti-crossing due to the interlayer hybridisation of excitons has also been found in reflectivity spectra of both $WS_2/MoS_2$ and $MoSe_2/WSe_2$ bilayers[12], which involved the B exciton in $WX_2$ and a layer-indirect exciton IX with electron on the tungsten and hole on the molybdenum layer. Here, we employ resonance Raman scattering (complemented by reflectivity and photoluminescence spectroscopy, all at 4 K) both to identify the anti-crossing features related to the IX$^*$ and B excitons hybridization in

WSe$_2$/MoSe$_2$ heterobilayers and to determine the layer composition of the superposition states of these excitons, together with the value of the interlayer coupling.

Raman scattering has a unique ability to determine the nature of the phonons responsible for scattering specific optically active electronic excitations, e.g. excitons. In addition, resonant enhancement of the Raman scattering when the incoming or outgoing photon energies coincide with specific exciton states means that resonance Raman scattering studies (rRs) permit us to access the same information on the exciton energies and linewidths as optical absorption and photoluminescence measurements. By observing the resonance behaviour of phonons due to both the WSe$_2$ and MoSe$_2$ monolayers[13,14] at hybridised excitons[8] in heterostructures one can directly quantify the exciton's layer composition. Below, we report the results of such studies performed on several MoSe$_2$/WSe$_2$ heterobilayers. These include sample HS1, where anti-parallel alignment of the 2D crystals (at $\theta=57°$) promotes the inter-intralayer exciton hybridization, and a reference sample HS2 ($\theta=20°$), where no interlayer hybridisation is expected due to a large momentum mismatch of the band edges in the two layers (additional samples with $\theta=6°$ and $\theta=60°$ are discussed in the supplementary information (SI)). The rRs studies were performed with excitation photon energies from 1.6 eV to 2.27 eV, enabling us to probe the excitonic resonances A1s, A2s, B1s and B2s in MoSe$_2$ and A1s, A2s and B1s and WSe$_2$ (A1s and B1s are the ground-state excitons in the two spin-orbit-split bands, and A2s and B2s are their first optically active excited states).

In Fig 1b we present Raman spectra for monolayers of WSe$_2$ and MoSe$_2$ when resonant with their respective A1s excitons along with Raman spectra for HS1 and HS2 measured in resonance with the WSe$_2$ intralayer A1s exciton. As expected, the Raman spectra for HS1 and HS2 are almost identical to the WSe$_2$ monolayer spectra with only minor variations in the relative intensity of different peaks. In Fig 1c we present Raman spectra for both monolayers and heterostructures taken with an excitation photon energy of 2.161 eV, which is near resonance with the WSe$_2$ B1s exciton. At this excitation energy, in sample HS1, we observe three new peaks at 291, 309 and 354 cm$^{-1}$ which do not appear in the other samples or at other resonances of HS1. In addition, in HS1 there is a peak at 241 cm$^{-1}$, which is anomalously strong (~ 3 times greater than in monolayer WSe$_2$ and HS2). As shown in Fig 2, these four peaks have a common resonance behaviour which is different from the other peaks observed in HS1 or the other samples (see SI Figures S5 & S6). Based upon Raman spectra reported for few layer WSe$_2$ and MoSe$_2$ [3,4], along with spectra for heterobilayers containing either MoSe$_2$ or WSe$_2$ layers[19–22] we assign the peaks at 309 and 354 cm$^{-1}$ to the WSe$_2$ and MoSe$_2$ A$_2''$(Γ) phonons. Based upon the assignment of a MoSe$_2$ phonon we also assign the peak at 291 cm$^{-1}$ to the MoSe$_2$ E′(Γ) phonon and the peak at 241 cm$^{-1}$ to the MoSe$_2$ A$_1'$(Γ). Both of which are observed at the same shift in the monolayer and heterostructure samples when resonantly exciting the MoSe$_2$ A1s and B1s excitons.

Whilst the observation of three MoSe$_2$ phonons at energies far from MoSe$_2$ resonances and near a predominantly WSe$_2$ B1s resonance is quite unexpected, it can be explained if the rRs probes an exciton which is a superposition of two excitons which involves electron/hole states in both the WSe$_2$ and MoSe$_2$ layers. We can also discard the possibility that these new peaks are hybridised phonon modes[19] as they are not observed at any of the other resonances. While the observed Raman scattering at MoSe$_2$ A1s and B1s and WSe$_2$ A1s resonances are only weakly perturbed by the other layer in the heterostructures, the WSe$_2$ B1s state is significantly hybridised with an exciton involving electrons in the MoSe$_2$ layer which excites MoSe$_2$ phonons, leading to the observation of peaks in the Raman spectra due to three of the four zone-centre MoSe$_2$ phonons. The fourth zone-centre phonon, the E″(Γ), is not observed in any spectra as it is forbidden by an in-plane symmetry which is likely inherited by the heterostructure[18,23].

With clear evidence for exciton hybridisation, we now determine the composition of the hybrid states. Without any MoSe$_2$ intralayer exciton near the WSe$_2$ B1s energy (see SI), the only other option is its hybridisation with an interlayer exciton. The energy of the fundamental interlayer exciton (IX), in which the electron is in MoSe$_2$ and the hole in WSe$_2$, is known[14,24] to have an energy of ~ 1.33 eV, and so we can discount it as the hybridised interlayer exciton. However, a higher-energy interlayer exciton (IX$^*$), which involves an electron in the WSe$_2$ layer and a hole in MoSe$_2$ layer (see Fig 3a) may have sufficient energy to hybridise resonantly with the B1s in WSe$_2$ via spin-conserving interlayer tunnelling of a hole. Using a combination of measured transition energies for HS1, experimental lattice and band structure parameters [25–29], and theoretical predictions for monolayer dielectric constants, interlayer band alignment and exciton binding energies [1,30–33](see Methods), we conclude that the IX* and WSe$_2$ B1s excitons are detuned by less than 220 meV. At twist angles other than 0 or 60°, there is a momentum mismatch between the bright (zero-momentum) intralayer B exciton and the bottom of the interlayer exciton dispersion. This leads to the energy of the interlayer exciton involved in the hybridisation increasing with twist angle, bringing it into resonance with the WSe$_2$ B1s exciton at $\theta \approx 51° \pm 9°$ (see SI).

Additional information, supporting the picture with the B1s and IX* hybridisation, comes from the fine structure of the rRs profiles of the Raman peaks measured in HS1, HS2 and the WSe$_2$ monolayer; key profiles are presented in Fig 4 with all the profiles and further discussion presented in the SI. The resonance behaviours of the monolayer and HS2 peaks are dominated by a single resonance (Fig 4a) at 2.156 ± 0.003 eV and 2.161 ± 0.006 eV respectively; the energies and errors were obtained by fitting the resonance behaviour to the standard third order perturbation theory (see SI). However, HS1 has clear evidence for two hybridised excitonic states in the same energy range, at 2.102 ± 0.001 eV and 2.168 ± 0.001 eV (Fig 4a). In HS1 the peaks also observed in WSe$_2$ monolayer show two resonances, one centred on the energy of each hybridised states (Fig 4a). However, the hybridisation specific peaks at 291, 309 and 354 cm$^{-1}$ show a single narrower resonance between the two energies (Fig 4b). This behaviour occurs because for these phonons scattering between the two hybridised excitons is stronger than scattering within each exciton band.

As observed in both the resonance Raman and reflectivity spectra, the lower energy hybridised state has a significantly lower oscillator strength $I_-$ than the upper's $I_+$, by a factor of $I_+/I_- \approx 11$ (see SI). As the oscillator strength of the non-hybridised interlayer exciton is negligible compared to the intralayer exciton, the oscillator strength ratio is a direct measurement of the intralayer exciton component of the two hybridised states (see SI). If we constrain the energies of the hybridised states to the measured values, we can then calculate their oscillator strength ratio and the non-hybridised states' detuning as a function of the tunnelling strength; Fig 3b and c. Based upon the measured oscillator strength ratio we determine the tunnel coupling as $t_v \approx 11$ meV and the energies of the non-hybridised B and interlayer excitons as 2.156 eV (in agreement with the monolayer B1s exciton) and 2.110 eV, respectively.

Besides providing direct experimental evidence for the hybridisation of layer-direct and -indirect excitons, resonant Raman scattering analysis of intra-layer phonons provides a method for a non-invasive measurement of the interlayer tunnelling for the spin and valley matching states in the crystallographically aligned layers. Obtaining quantitative information on the interlayer hybridisation constants for the states in heterostructures of two-dimensional semiconductors opens new opportunities for the intelligent design of heterostructures for optoelectronics applications in the areas of quantum technologies. The hybridised layer-direct and -indirect excitons represent a two-level system, which is tunable by an externally applied vertical displacement field [8], and where the superposition of two modes can be operated as an optical qubit. Placing such bilayers inside an optical cavity with a localised photon mode close to the hybridised exciton pair [10,34–36] may additionally open ways towards making a three-level quantum-optical bit for ultra-fast information processing.

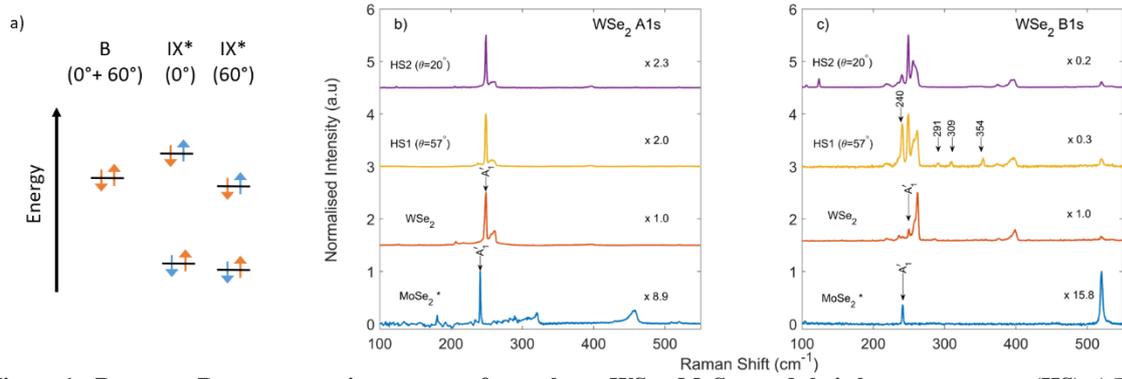

**Figure 1** - **Resonant Raman scattering spectra of monolayer WSe$_2$, MoSe$_2$, and their heterostructures (HS). a)** Energy level diagram for the WSe$_2$ B exciton and the interlayer excitons which can couple to it by spin-conserving single-particle tunnelling. The arrows represent the spin of the electron and hole respectively and the colour represents the material; orange for WSe$_2$ and blue for MoSe$_2$. The relative energies have been estimated using measured optical transition energies and the theoretical predictions for the spin-orbit couplings. **b)** Raman spectra for monolayer WSe$_2$, HS1 (57°) and HS2 (20°) when resonant with the WSe$_2$ A1s exciton and a spectra for a monolayer MoSe$_2$ when resonant with the MoSe$_2$ A1s exciton. **c)** Raman spectra for an excitation energy 2.161 eV near resonance with the WSe$_2$ B1s exciton. To allow for ease of visual comparison each spectra has been normalised to the maximum intensity and offset. The scaling factors shown in each panel are relative to the absolute scattering probability of the WSe$_2$ A$_1'$ peak. In panel b) the arrows highlight the new peaks at 241, 291, 309 and 354 cm$^{-1}$.

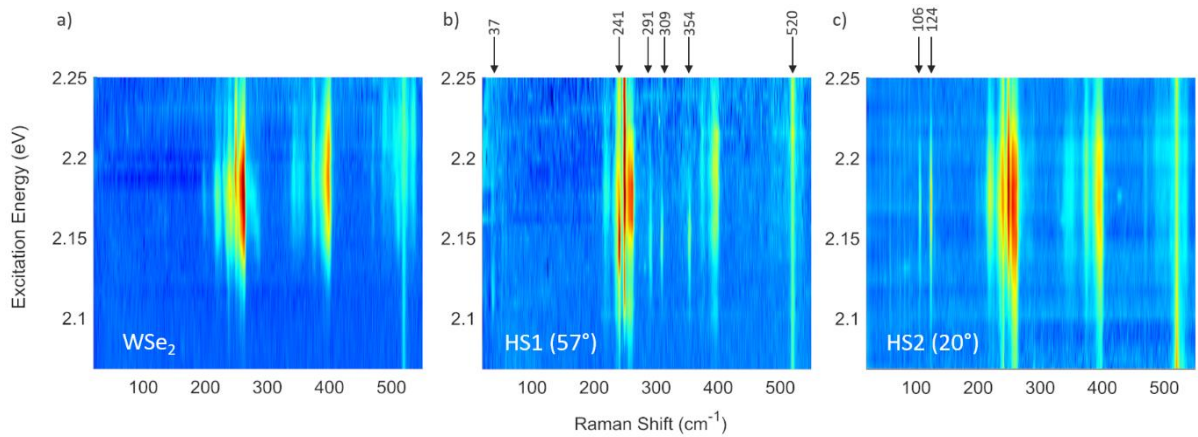

**Figure 2** - **Comparison of Raman intensity plots between monolayer and heterostructures.** Resonance Raman spectra for **a**) monolayer WSe$_2$, **b**) HS1 (57°), and **c**), HS2 (20°) are presented for excitation energies from 2.05 to 2.25 eV. The data is plotted using a logarithmic scale for the intensity in order to highlight weak features. Arrows included along the top of panels b) & c) indicate the position of the new peaks in HS1 (241, 291, 309 and 354 cm$^{-1}$), two defect activated peaks in HS2 106 and 124 cm$^{-1}$, along with the Si Raman peak at 520 cm$^{-1}$. The peak at 37 cm$^{-1}$ in HS1 is attributed to a moire zone-folded acoustic phonon like those reported in homobilayers[37]. For monolayer WSe$_2$ all the peaks appear to have a broad resonance centred at ~ 2.180 eV. Likewise HS2 all the peaks have a broad resonance centred at ~2.170 eV. In contrast in HS1 whilst the WSe$_2$ peaks at 257, 261 and 396 cm$^{-1}$ have a resonance centred at ~ 2.18, the new peaks at 241, 291, 309 and 354 cm$^{-1}$ have a resonance which peaks at a lower energy ~2.15 eV and is narrower.

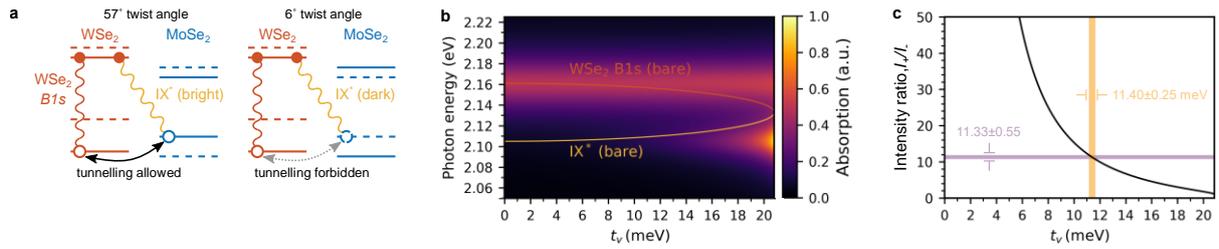

**Figure 3 - Twist angle dependent hybridisation of the WSe$_2$ B1s and bright IX$^*$ interlayer excitons. a)** Schematic of the two excitons, where electrons (holes) are represented by filled (empty) circles; note the energies represented on the vertical scale are single particle energies which are modified by the excitonic binding energies. Solid and dashed lines represent valley states of opposite spin projection in the two TMDs. In a 57° heterostructure, the proposed interlayer exciton IX$^*$, which is close to resonance with WSe$_2$ B1s, is bright and can couple to the latter through interlayer hole tunnelling. By contrast, in a 6° structure, the spins of interest in MoSe$_2$ and WSe$_2$ are oppositely orientated. IX$^*$ is thus dark and forbidden from hybridizing with WSe$_2$ B1s. **b)** Calculated optical absorption spectrum of hybridized excitons (normalized) for a 57° heterostructure, as a function of the hole tunnelling energy $t_v$. The bare WSe$_2$ B1s and IX$^*$ energies corresponding to each value of $t_v$ are shown with solid lines. **c)** Absorption intensity ratio between the higher and lower energy absorption lines. The experimentally measured value is shown in purple and the range of $t_v$ consistent with this measurement is highlighted in yellow. This places the detuning between WSe$_2$ B1s and IX$^*$ at $46.9 \pm 0.4$ meV. See Methods for calculation details.

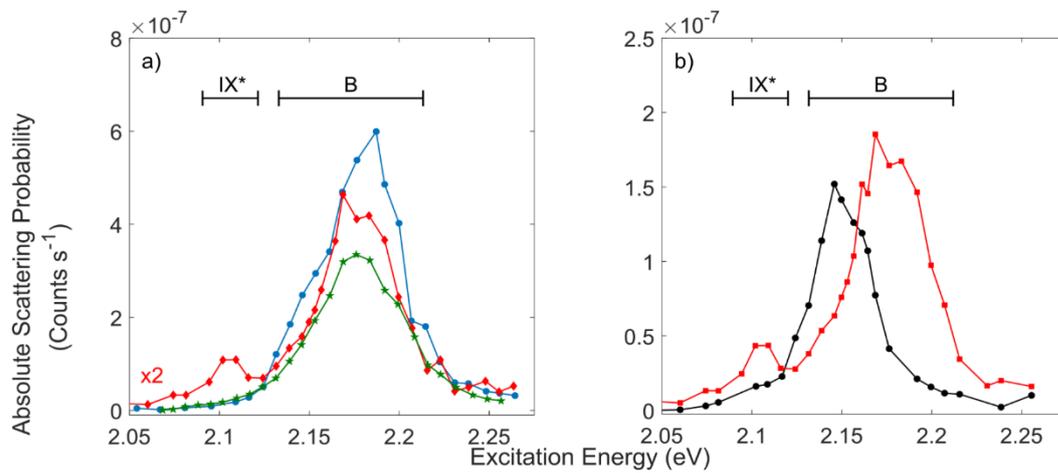

**Figure 4 - Resonance Raman profiles for key Raman peaks. a)** Resonance profiles for the 354 cm$^{-1}$ peak in the WSe$_2$ monolayer (blue), HS1 (θ=57°) (red) and HS2 (θ=20°) (green). **b)** Comparison of the resonance profiles of the 354 cm$^{-1}$ peak, as a representative of the monolayer observed peaks, and 257 cm$^{-1}$, as a representative of the hybridisation allowed peaks, of HS1. The B exciton/upper hybridised state and the lower hybridised state IX$^*$ regions are marked.

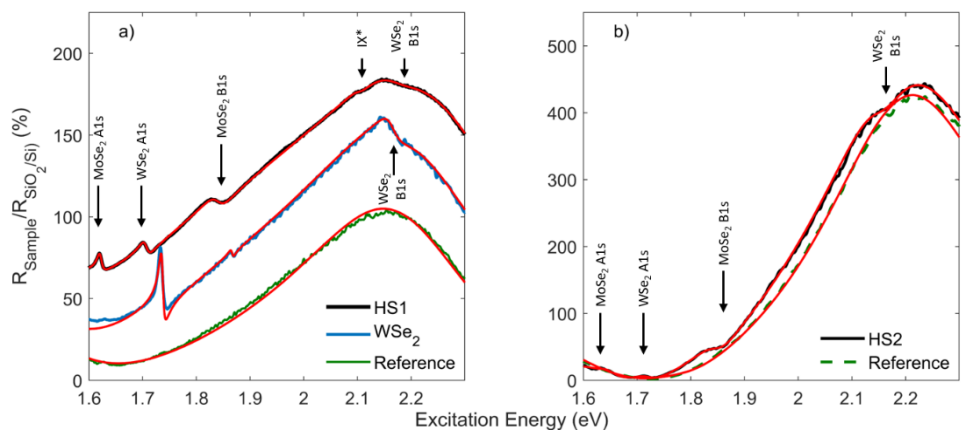

**Figure 5 - Reflectivity spectra for both heterobilayer and monolayer regions.** The reflectivity spectra measured relative to the SiO$_2$/Si substrate are presented for **a)** both HS1 and monolayer WSe$_2$ and **b)** HS2 alongside reference spectra obtained on comparable regions without the TMD layers. In panel a) the reflectivity spectra for monolayer WSe$_2$ and HS1 have been offset from the reference spectra by 30% and 60% respectively to allow for easier visual comparison. In each case the red lines

indicate fits to the reflectivity spectra performed using T-Matrix calculations; the model used to fit the TMD structures differs only from that used for the reference due to the contributions of the TMD layers. A complete set of fit parameters is given in the SI and the energies of key transitions are discussed in the main body of the paper.

# Methods

**Experimental Methods**

The heterobilayers used in our experiments consist of mechanically exfoliated monolayers of MoSe$_2$ and WSe$_2$ encapsulated between layers of hexagonal boron nitride using a dry transfer technique [38]. The underlying substrate is oxide coated silicon. For all measurements the samples were mounted inside an Oxford Instruments High Resolution liquid helium flow microstat, and cooled to 4 K. All optical measurements were carried out using a back-scattering geometry with a x50 Olympus objective resulting in a spot size of ~ 3 μm. Positioning on the sample was achieved using a computer controlled 3 axis translation stage and a custom in situ microscope. For all measurements the incident power on the sample was maintained below 100 μW to avoid photo doping and laser heating [39,40]. The resonance Raman measurements were carried out using a CW Coherent Mira 900 allowing excitation energies from 1.24 to 1.77 eV and a Coherent Cr 599 dye laser using DCM, Rhodamine 6G and Rhodamine 110 laser dyes allowing excitation energies from 1.74 to 2.27 eV. The incident polarizations of the lasers were horizontal relative to the optical bench and the Raman scattered light coupled into the spectrometer was analysed using both horizontal and vertical polarizations. The polarization of the Raman peaks was observed to be strongly co-linear. This allows unwanted luminescence from the samples to be removed from the Raman spectra by subtraction of the crossed and parallel polarized spectra. The Raman spectra were recorded using a Princeton Instruments Tri-vista Triple spectrometer equipped with a liquid nitrogen cooled CCD. The Raman peak frequencies were all calibrated using the silicon peak at 520 cm$^{-1}$ as an internal reference. To allow comparison of the Raman scattering intensity across the different samples the spectra were calibrated to absolute Raman scattering probability. This required the normalization of the Raman spectra to the 520 cm$^{-1}$ silicon peak intensity, correction of Fabry-Perot interference effects and calibration using the absolute Raman scattering results[16]. To account for the Fabry-Perot interference we made use of reflectivity spectra measured using a Fianium super continuum source and Ocean optics HR4000 spectrometer.

For further detail on the experimental methods, and data analysis please refer to the supplementary information and our previous work exploring resonance Raman scattering in monolayer TMDs.

**Theoretical Model**

**Estimation of the WSe$_2$ B1s and interlayer exciton IX$^*$ detuning in anti-parallel stacked WSe$_2$/MoSe$_2$.** To approximate the IX$^*$ energy we solved the Wannier equation for a MoSe$_2$ hole and a WSe$_2$ electron with effective masses extracted from the earlier publications [25–29] and using the long-range interlayer electron-hole interaction derived in Ref [41]. The layer-indirect gap between MoSe$_2$ valence-band and WSe$_2$ conduction-band edges (see Fig. 3a), was estimated from the ab initio DFT data reported in Refs. [33,42] and experimental and DFT spin-orbit splitting energies for the MoSe$_2$ valence band and WSe$_2$ conduction band, respectively [1,29]. Combining this with the exciton binding energy gives a total energy of $2.039 \pm 0.111$ eV for the momentum-indirect IX$^*$ state, and we estimate an additional kinetic energy of $\approx 17$ meV for the bright (momentum-direct) IX$^*$ due to momentum mismatch determined by the interlayer twist angle. Contrasting this with the measured WSe$_2$ B1s energy of $2.164 \pm 0.001$ eV we conclude that the detuning between the two excitons is between 3 meV and 219 meV.

**Estimation of the interlayer valence-band hopping energy from the resonance Raman spectrum.** Hybridization between the intralayer WSe$_2$ B1s and interlayer IX$^*$ excitons can be modelled by a $4 \times 4$ Hamiltonian with one basis state corresponding to WSe$_2$ B1s, and three representing the valley-mismatched interlayer excitons formed by holes at the three equivalent K valleys of MoSe$_2$, each bound to an electron at the nearest K valley of WSe$_2$. The intralayer-interlayer exciton hybridization term is related to the interlayer valence-band hopping energy $t_v$, defined in Ref. [8] and SI. Diagonalizing the Hamiltonian gives two bright states $\psi_-$ and $\psi_+$ at energies $E_-$ and $E_+$ with oscillator strengths deriving from their WSe$_2$ B1s exciton components, and two dark states formed solely by interlayer excitons, which we shall ignore henceforth. We computed the oscillator strength ratio of the two bright states as the modulus squared of the ratio of their WSe$_2$ B1s components for a range of values of $t_v$, fixing the energies $E_-$ and $E_+$ at the experimental values for the hybridised exciton resonances, as determined by the rRS data in Fig. 3b. Comparing the theoretical oscillator strength ratios to that measured by resonance Raman gives a hopping strength $t_v \approx 11$ meV. The available data on the interlayer tunnel coupling for electrons are between 26 and 43 meV [11,12], comparable with our obtained value.

## Data Availability

The data presented in this paper is openly available from the University of Southampton Repository at DOI: 10.5258/SOTON.D1314.

## Author Contributions

The experiments were conceived by D.C.S, L.P.M and X.X. Samples were fabricated by P.R. The experimental measurements were performed by J.V and L.P.M. Theoretical calculations were performed by D.R.T and V.F. Data analysis and interpretation was carried out by L.P.M, D.C.S, J.V, D.R.T and V.F. The paper was written by D.C.S, L.P.M, D.R.T and V.F. All authors discussed the results and commented on the manuscript.

## Corresponding Author

*D.C.Smith@soton.ac.uk

Present Addresses

## Competing financial interests

The authors declare no competing financial interests.

## Funding Sources

This research was supported by UK Engineering and Physical Sciences Research Council via program grant EP/N035437/1. Both L.P.M and J.V were supported by EPSRC DTP funding. The work at U.

Washington was funded by the Department of Energy, Basic Energy Sciences, Materials Sciences and Engineering Division (DE-SC0018171). D.R-T. was funded by UNAM-DGAPA.

# Supplementary Information for Superposition of intra- and inter-layer excitons in twistronic MoSe$_2$/WSe$_2$ bilayers probed by resonant Raman scattering

*Liam P. McDonnell[1], Jacob J.S. Viner[1], David A. Ruiz-Tijerina[2], Pasqual Rivera[3], Xiaodong Xu[3], Vladimir I. Fal'ko[4,5] and David C. Smith[1*]*

[1] School of Physics and Astronomy, University of Southampton, Southampton SO17 1BJ, United Kingdom.

[2] Centro de Nanociencias y Nanotecnología, Universidad Nacional Autónoma de México, C.P. 22800, Ensenada, Baja California, México.

[3] Department of Physics, University of Washington, Seattle, WA, USA.

[4] National Graphene Institute, University of Manchester, M13 9PL, United Kingdom.

[5] Henry Royce Institute for Advanced Materials, University of Manchester, Manchester, M13 9PL, United Kingdom.

*Corresponding author – D.C.Smith@soton.ac.uk

## Contents

# Additional Raman Spectra

The Raman spectra shown in Figures S1 & 2 present additional Raman spectra for two other heterobilayers with twist angles of 6 and 60°, when resonant with the WSe$_2$ A1s and B1s excitons. In Figure S1 there is no evidence of the new peaks observed on HS1 at 291, 309 or 354 cm$^{-1}$ in either the 6 or 60 ° heterobilayers. On HS4 we also observe the same defect-activated peaks that are seen in HS2 at 106 and 124 cm$^{-1}$.

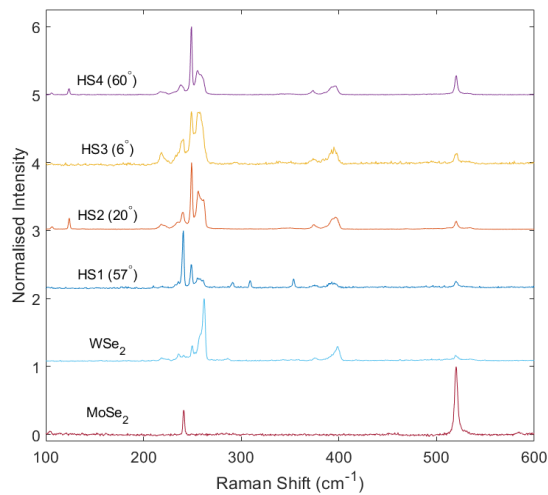

***Figure S1*** *Raman spectra for a WSe$_2$ monolayer and four MoSe$_2$/WSe$_2$ heterobilayer (θ=57, 20, 6, 60). Spectra were obtained with an excitation energy of 2.156 eV at 4 K. To allow for a comparison of the spectra the Raman peaks have been normalised to the most intense peak in the spectra and offset. For HS1 we observe three new peaks which are not observed on any of the other heterobilayers or in monolayer WSe$_2$ at 291, 309 and 354 cm$^{-1}$.*

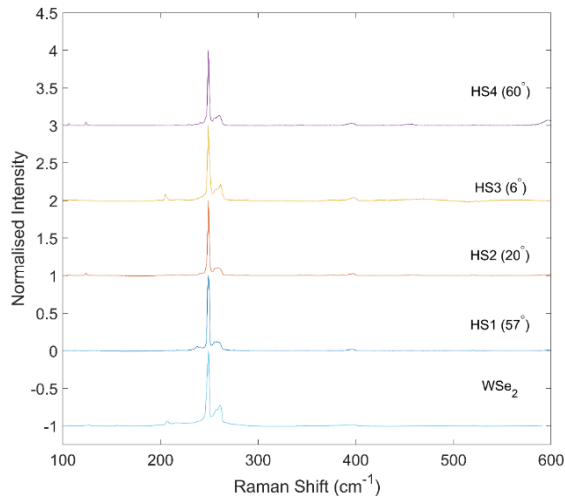

***Figure S2*** *Raman Spectra for monolayer WSe$_2$ and the four MoSe$_2$/WSe$_2$ heterobilayers (θ=57, 20, 6, 60) when resonant with the WSe$_2$ A1s exciton. For ease of comparison all spectra have been normalised to the intensity of the WSe$_2$ A$_1$'/E' peak at 249 cm$^{-1}$ and offset.*

# Monolayer MoSe$_2$ Resonance Raman Data

Figure S3 presents resonance Raman data for a MoSe$_2$ monolayer with excitation energies between 1.8 and 2.27 eV. The B1s and B2s excitons with both their incoming and outgoing resonances are visible below 2.1 eV. It is clear there is no additional MoSe$_2$ intralayer exciton between 2.1 and 2.27 eV. At these excitation energies the MoSe$_2$ Raman spectra are non-resonant with only the A$_1'$ Raman peak visible and, as discussed in the main body of the paper, is an order of magnitude weaker than the WSe$_2$ Raman signal. Therefore, the hybridised state observed in HS1 must involve the WSe$_2$ B1s and an interlayer exciton.

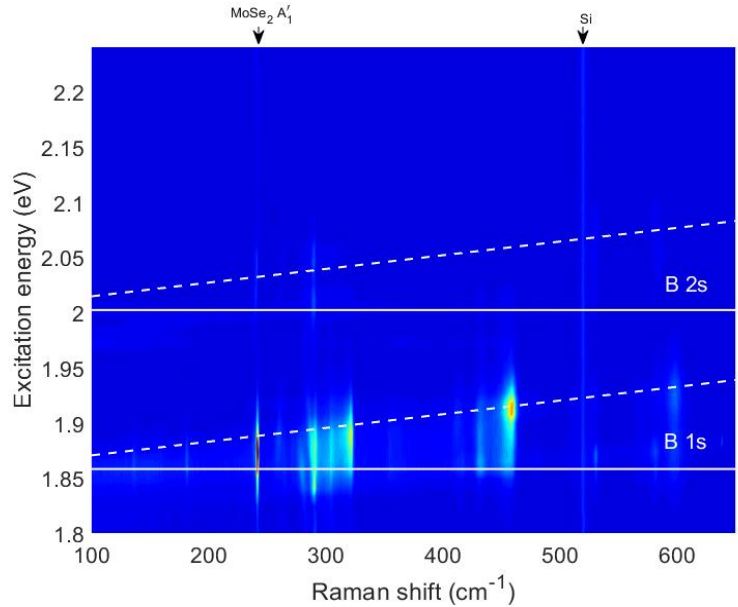

*Figure S3* Colour plot showing resonance Raman data for monolayer MoSe$_2$ from 2.07 to 2.24 eV. We observe the B1s and B2s resonances below 2.09 eV, where both the incoming (solid white line) and outgoing resonances (dashed white line) are indicated. At higher excitation energies the Raman spectra are non-resonant with only the MoSe$_2$ A$_1'$ and 520 cm$^{-1}$ silicon peaks visible. Between 2.05 and 2.25 eV we observe no additional resonance enhancement of the MoSe$_2$ phonon. This demonstrates that there are no additional intralayer MoSe$_2$ excitons at the energies expected for the WSe$_2$ B1s exciton.

# Fitted Reflectivity Spectra

The reflectivity spectra presented in the main body of the paper were measured relative to the $SiO_2$/Si substrate and fitted using a T-matrix calculation. The refractive indices used for Graphite, hBN, $SiO_2$ and Si are those reported in ref [S1–4]. The TMD layers were fitted using a dielectric function of the form given in Eq 1 to determine their refractive indices at 4 K. Where $E_k$ and $\Gamma_k$ are the energy and broadening parameter of the excitonic state and $a_k$ is the amplitude of the Lorentzian. To constrain the thicknesses of the layers in the Van der Waals structure a series of spectra were obtained on areas with different stacking and fitted using our T matrix model.

$$\epsilon(E) = 1 + \sum_{k=1}^{N} \frac{a_k}{E_k^2 - E^2 - i2E\Gamma_k}$$

$a_k = \frac{\sum_i \left[\langle i|\hat{d}|0\rangle\right]^2 E_k}{\varepsilon_0 V}$ ,where the summation is over all states at the energy $E_k$  (1)

Fitting the reflectivity spectra for the different heterobilayers allows us to determine the oscillator strength, with and energies of the excitonic states as shown in Table S1.

| Sample | Exciton | Amp | $\Gamma$ (meV) | E (eV) |
|---|---|---|---|---|
| WSe$_2$ | WSe$_2$ A1s | 1.19 ± 0.02 | 3.3 ± 0.1 | 1.737 ± 0.001 |
|  | WSe$_2$ B1s | 0.72 ± 0.04 | 20.2 ± 1.1 | 2.164 ± 0.001 |
| HS1 | MoSe$_2$ A1s | 0.73 ± 0.01 | 5.0 ± 0.1 | 1.622 ± 0.001 |
|  | MoSe$_2$ B1s | 0.91 ± 0.01 | 18.7 ± 0.1 | 1.841 ± 0.001 |
|  | WSe$_2$ A1s | 0.68 ± 0.01 | 8.2 ± 0.1 | 1.707 ± 0.001 |
|  | hX1 | 0.1 ± 0.01 | 13.6 ± 0.6 | 2.102 ± 0.001 |
|  | hX2 | 1.08 ± 0.01 | 43.6 ± 0.4 | 2.168 ± 0.001 |
| HS2 | MoSe$_2$ A1s | 1.24 ± 0.19 | 16.5 ± 2.0 | 1.640 ± 0.001 |
|  | MoSe$_2$ B1s | 1.07 ± 0.04 | 25.46 ± 0.89 | 1.851 ± 0.001 |
|  | WSe$_2$ A1s | 1.13 ± 0.07 | 15.35 ± 1.5 | 1.720 ± 0.001 |
|  | WSe$_2$ B1s | 0.64 ± 0.02 | 31.06 ± 0.96 | 2.161 ± 0.001 |

*Table S1 Coefficients from fitting reflectivity spectra for monolayer WSe$_2$, HS1 and HS2 using T matrix methods. Errors given for coefficients are a standard deviation obtained from the fitting process.*

# Resonance Raman Profiles

## Absolute Raman Scattering Cross Section

All resonance Raman data presented has been calibrated to give the absolute Raman scattering probability of the TMD peaks. This was achieved by correcting for spectrometer throughput followed by normalisation of the data to the silicon peak at 520 cm$^{-1}$. The spectra were then corrected using a Raman enhancement factor to correct for Fabry-Perot effects [S5]. Finally the data was calibrated using the absolute scattering efficiency for silicon obtained by Aggarwal et al. [S6] who measured the absolute scattering cross section of the 520 cm$^{-1}$ Raman peak.

When accounting for the Fabry Perot effects we calculated a Raman enhancement factor using the approach described by Yoon et al. [S5] and using the TMD refractive indices determined from fitting the reflectivity spectra. The Raman enhancement factor has the form given in Eq 2. Where $E_{Ab}$ and $E_{Sc}$ are the electric field strengths at the absorbed and scattered photon energies, x is the depth in the material from the interface and d denotes the thickness of the layer.

$$\int_0^d |E_{Ab}(x) E_{Sc}(x)|^2 \, dx \qquad (2)$$

## Resonance Raman scattering models

The resonance Raman profiles presented in the paper and in the following sections have been fitted using the standard Raman scattering model derived from third order perturbation theory. The expression for the absolute Raman scattering probability derived from this model is given in Eq 3.

$$P_{Raman} = \frac{1}{\pi \hbar^4 c^5 \epsilon_0^2} \left| \sum_{i,j} \frac{\underline{\sigma}_{sc} \cdot \langle +1|\hat{d}|j\rangle \langle j|\hat{H}_{ex-ph}|i\rangle \langle i|\hat{d}|0\rangle \cdot \underline{\sigma}_{in}}{(E - E_i - i\Gamma_l)(E - E_j - E_{p1} - i\Gamma_j)} \right|^2 E_{in} E_{sc}^3 \qquad (3)$$

Where $\hat{d}$ is the dipole operator and $\underline{\sigma}_{sc}$ and $\underline{\sigma}_{in}$ are unit polarization vectors for the incoming and scattered photons (energies $E_{in}$ and $E_{sc}$ respectively), $\hat{H}_{ex-ph}$ is the exciton-phonon interaction Hamiltonian, $|i\rangle$ and $|j\rangle$ are intermediate excitonic states of the system, $|0\rangle$ is the initial state of a system and $|+1\rangle$ is the final state of the system with one additional phonon.

## Monolayer WSe₂ Resonance Profiles

For monolayer WSe$_2$ it is possible to fit almost all the resonance Raman profiles using a Raman scattering model with a single underlying excitonic state. The only exception is the 249.1 cm$^{-1}$ peak which has an additional shoulder at ~2.15 eV. This can be explained by fitting with an additional excitonic state at ~ 2.144 eV. The exact nature of this state, as discussed in the main body of the paper, is unclear and it is not observed in reflectivity spectra. Excluding the behaviour of the 249.1 cm$^{-1}$ peak we find that a single state can well describe the other resonance profiles and results in a mean energy and width for the underlying excitonic state of 2.156 ± 0.002 eV and 23.2 ± 2.6 meV. The fitted resonance Raman profiles and obtained coefficients are shown in Fig S4 and Table S2.

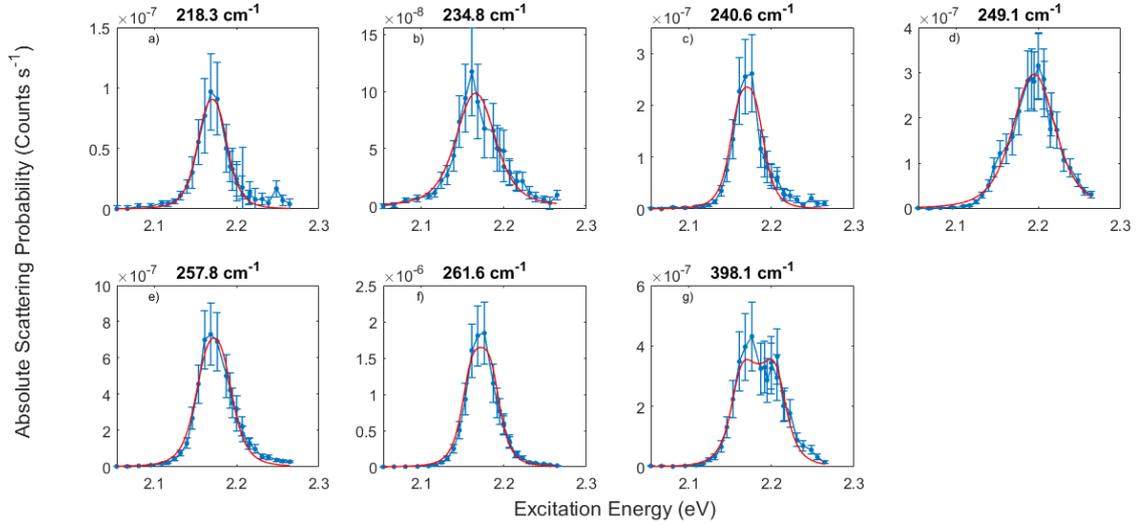

***Figure S4*** *Resonance Raman profiles obtained on monolayer WSe$_2$ when resonant with the B1s exciton. All resonance profiles have been fitted to a single excitonic state model and the resultant fits are shown by the red lines. Error bars shown are a standard deviation obtained from the fitting process.*

| Raman Shift (cm$^{-1}$) | $\|A_1\|^2$ (eV$^4$) | Width (meV) | Energy (eV) |
|---|---|---|---|
| **218.3** | 0.4±0.1 | 21.1±1.7 | 2.157±0.001 |
| **234.8** | 2.8±0.5 | 36.6±2.6 | 2.151±0.001 |
| **240.6** | 0.6±0.1 | 18.9±2.5 | 2.156±0.001 |
| **249.1** | 11.6±1.8 | 40.4±1.9 | 2.179±0.001 |
| **257.8** | 3.7±0.8 | 23.6±1.6 | 2.156±0.001 |
| **261.6** | 5.6±0.9 | 18.5±1.9 | 2.156±0.001 |
| **398.1** | 3.0±0.6 | 20.2±2.1 | 2.160±0.001 |

***Table S2*** *Fitted coefficients for fitting monolayer WSe$_2$ Resonance profiles at the B1s resonance using a single excitonic state model. Amplitudes are scaled by a factor of 10$^{13}$. Errors given are a standard deviation obtained from the fitting process.*

# HS2 (20°) Resonance Profiles

As discussed in the main body of the paper the resonance profiles for the HS2 Raman are dominated by a single excitonic state. The resonance profiles were fitted using a Raman scattering model with a single underlying excitonic state. For the peaks at 239.9 and 249.2 cm$^{-1}$ there is also a resonance contribution from a state above 2.27 eV which is not resolved in our experiments and is likely the C exciton. As a result, these profiles require a two-state fit including an additional state whose energy and width are fixed to 2.3 eV and 50 meV, respectively. The results of fitting are shown in Figure S5 and the coefficients obtained are presented in Table S3. The widths and energies of the underlying transition are in reasonable agreement with a mean energy of 2.161 ± 0.006 eV and width of 38.7 ± 6.3 meV. Therefore, for HS2 ($\theta$=20°) the resonance behaviour is dominated by a single state at 2.161 eV and attributed to the WSe$_2$ B1s exciton.

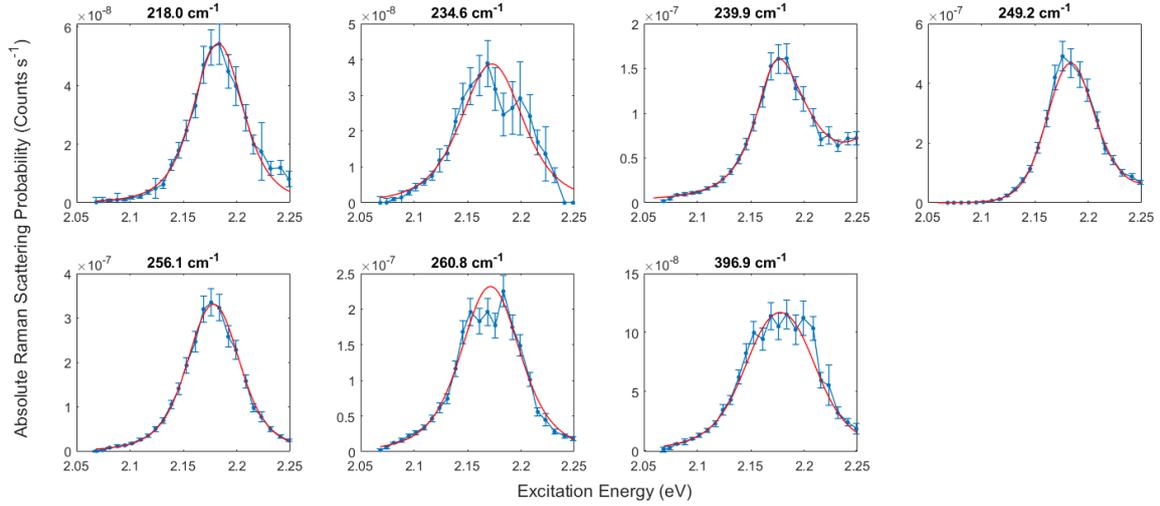

**Figure S5** Resonance Raman profiles are presented for HS2 (20°) when resonant with the WSe$_2$ B1s exciton and the resultant fits are indicated by the red lines. The resonance profiles for the peaks at 218, 236, 257, 261 and 396 cm$^{-1}$ have all been fitted using a single excitonic state. For the peaks at 240 and 249 cm$^{-1}$ the resonance profiles clearly contain contributions from an unresolved state above 2.27 eV, and so were fitted using a two state model with an unbound lower energy state between 2.1 and 2.2 eV and with a higher energy state fixed at 2.3 eV. For all profiles, the error bars shown are a standard deviation determined from the fitting process.

| Raman Shift (cm$^{-1}$) | $\|A_1\|^2$ (eV$^4$) | Width (meV) | Energy (eV) |
|---|---|---|---|
| 218.0 | 1.26±0.11 | 36.6±1.0 | 2.169±0.001 |
| 234.6 | 2.48±0.68 | 48.1±4.0 | 2.158±0.002 |
| 239.9 | 1.76±0.35 | 24.4±6.9 | 2.170±0.006 |
| 249.2 | 5.45±1.77 | 28.8±11 | 2.164±0.001 |
| 256.1 | 8.69±0.39 | 37.0±0.5 | 2.162±0.001 |
| 260.8 | 10.99±1.39 | 43.8±1.7 | 2.155±0.001 |
| 396.9 | 6.82±1.01 | 42.5±2.2 | 2.153±0.001 |

**Table S3** Fitted Coefficients from fitting resonance profiles on HS2 (20°) shown in Fig S5. The coefficients for the peaks at 240.2 and 249.0 cm$^{-1}$ are shaded in grey to indicate that these were obtained using a two state model with a higher energy state fixed at 2.3 eV scattering amplitudes with this higher state are 484.44±27.4 and 910.23±34.97. Amplitudes given are scaled by a factor of 10$^{13}$. Errors given are a standard deviation obtained from the fitting process.

# HS1 (57°) Resonance Profiles: Unconstrained Analysis

As discussed in the main body of the paper for HS1 we require a minimum of two excitonic states to fit the resonance profiles. For the peaks at 240.5 and 249.1 cm$^{-1}$ an additional excitonic state is also required to account for the observed resonance at higher excitation energies from ~ 2.2 to 2.27 eV, which is attributed to C excitons. As a result, when fitting these peaks an additional state is included with fixed width and energy of 50 meV and 2.3 eV. The results of fitting the resonance profiles are shown in Figure S6 with the obtained coefficients presented in Table S4. The resulting energies and widths for the two excitonic states clearly indicate two transitions with widths and energies of 2.111 ± 0.005 eV and 27.2 ± 11.7 meV, and 2.162±0.008 eV and 28.7 ± 10.91 meV.

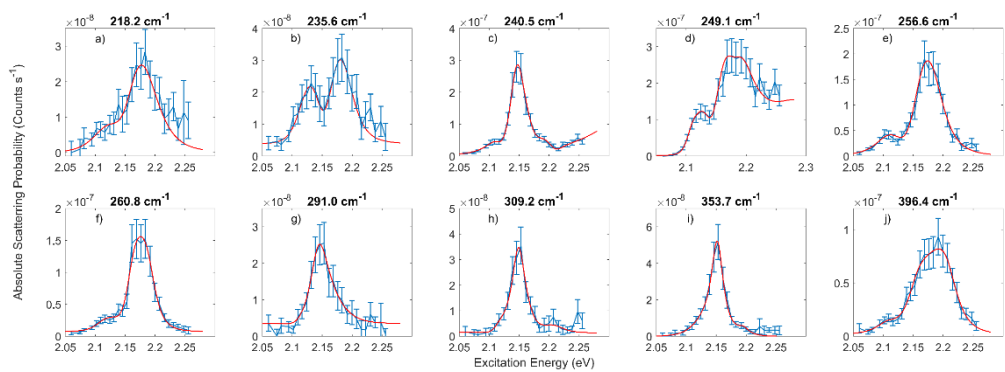

*Figure S6 Resonance Raman profiles are shown for HS1 when resonant with the hybridised excitonic states along with fits obtained using two state and three state models where appropriate. For this analysis all fitted profiles were obtained with the widths and energies of the excitonic states free to vary. Error bars shown are a standard deviation obtained from fitting the Raman spectra.*

| Raman Shift (cm$^{-1}$) | $\|A_1\|^2$ (eV$^4$) | $\|A_2\|^2$ (eV$^4$) | $\|A_3\|^2$ (eV$^4$) | $\Gamma_1$ (meV) | $\Gamma_2$ (meV) | $E_1$ (eV) | $E_2$ (eV) |
|---|---|---|---|---|---|---|---|
| **218.2** | 2.53±1.37 | 115.27±16.87 | 87.96±22.92 | 50.87±23.27 | 50.12±10.97 | 2.115±0.011 | 2.153±0.08 |
| **235.6** | 0.04±0.15 | 7.79±0.99 | 3.64±1.77 | 18.42±5.89 | 32.45±4.19 | 2.115±0.069 | 2.159±0.029 |
| **240.5** | 77.2±8.07 | 4.05±0.21 | 15.18±1.36 | 16.91±1.92 | 16.72±11.84 | 2.115±0.003 | 2.178±0.013 |
| **249.1** | 27.05±2.21 | 6.18±0.96 | 1.11±0.27 | 15.17±9.73 | 23.90±11.42 | 2.105±0.016 | 2.161±0.007 |
| **256.6** | 0.45±0.11 | 24.33±0.82 | 0.34±0.09 | 18.5±8.22 | 30.02±4.13 | 2.101±0.007 | 2.156±0.001 |
| **260.8** | 2.67±1.26 | 18.61±0.45 | 8.6±2.26 | 24.94±14.93 | 22.6±4.93 | 2.108±0.008 | 2.159±0.004 |
| **291.0** | 0.27±0.4 | 0.03±0.06 | 1.13±0.99 | 19.15±8.45 | 39.35±13.48 | 2.106±0.009 | 2.168±0.012 |
| **309.2** | 0.83±5.01 | 0.54±0.58 | 2.84±1.18 | 13.88±3.43 | 24.26±13.44 | 2.110±0.003 | 2.172±0.012 |
| **353.7** | 9.44±1.14 | 0.58±0.27 | 3.08±0.98 | 32.52±15.9 | 12.33±1.68 | 2.118±0.009 | 2.152±0.002 |
| **396.4** | 0.61±0.48 | 17.52±0.97 | 0.17±0.46 | 25.07±12.08 | 29.9±2.84 | 2.107±0.008 | 2.160±0.001 |

*Table S4* Energies and Widths obtained from fitting the resonance Raman profiles for HS1 using a two-state model. The exceptions to this are the peaks at 240.5 and 248.9 cm$^{-1}$ (shaded in gray) which demonstrate a significant resonant contribution at 2.27 eV and so are fitted using a three state model including an additional state fixed at 2.3 eV with a width of 50meV. The resulting amplitudes for this higher energy state are 360±51 and 578±99, for the 240.5 and 248.9 cm$^{-1}$ peaks respectively. In this instance the energies, widths and amplitudes of the scattering processes are not constrained. Amplitudes are scaled by a factor of $10^{14}$.

# HS1 (57°) Resonance Profiles: Constrained Analysis

In addition to the energies and linewidths of the underlying excitons our two state Raman scattering model contains three amplitudes associated with the different possible scattering channels. Where $A_1$ and $A_2$ give the scattering rates for each of the two excitonic states and $A_3$ is the interstate scattering rate between the two transitions. A comparison of these amplitudes provides further information on the underlying scattering processes. However, the resulting amplitudes obtained from our unconstrained analysis of HS1 are difficult to interpret due to the significant variation in the fitted exciton linewidths and energies. Therefore, we performed a constrained analysis where the energies and linewidths of the two transitions were fixed to the those obtained from fitting the WSe$_2$ $A_1'$ 249 cm$^{-1}$ peak. This peak was chosen as it is the only WSe$_2$ Raman peak which is directly attributed to a single phonon scattering process and so should be best described by our model. The resulting amplitudes obtained from our constrained analysis reveals that the WSe$_2$ Raman peaks are dominated by scattering with the higher energy excitonic state at 2.161 eV, which is closest to the energy of the WSe$_2$ B1s intralayer exciton. Whereas the dominant scattering channel associated with the new Raman peaks at 240.5, 291.0, 309.2 and 353.7 cm$^{-1}$ is due to interstate scattering.

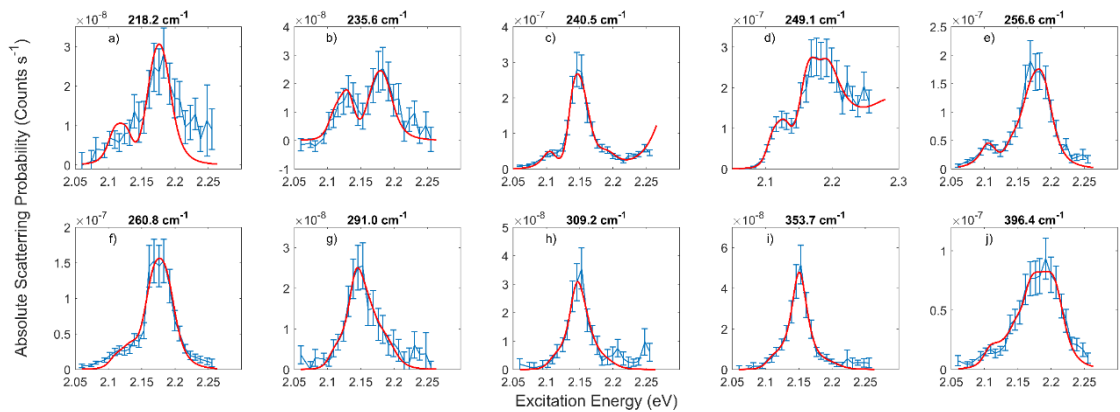

*Figure S7* Fitted Resonance profiles for HS1. For this analysis the energies and widths used in the fitting process have been constrained using the values obtained from fitting the WSe$_2$ $A_1'$ peak at 249.1 cm$^{-1}$. Error bars shown are a standard deviation obtained from the fitting process.

| Raman Shift (cm$^{-1}$) | $\|A_1\|^2$ (eV$^4$) | $\|A_2\|^2$ (eV$^4$) | $\|A_3\|^2$ (eV$^4$) |
|---|---|---|---|
| 218.2 | 0.15±0.25 | 2.08±0.31 | 0.4±0.85 |
| 235.6 | 0.02±0.1 | 1.98±0.15 | 1.33±0.89 |
| 240.5 | 9.02±2.1 | 0.38±0.25 | 11.99±2.95 |
| 249.1 | 7.37±0.81 | 28.27±0.73 | 4.15±0.8 |
| 256.6 | 2.17±0.55 | 14.6±1.33 | 0.61±0.43 |
| 260.8 | 0.85±0.57 | 12.69±0.83 | 1.33±1.3 |
| 291.0 | 0.01±0.01 | 0.22±0.07 | 0.35±0.05 |
| 309.2 | 0.02±0.02 | 0.12±0.06 | 0.32±0.04 |
| 353.7 | 1.96±0.87 | 0.3±0.18 | 3.14±0.6 |
| 396.4 | 0.82±0.62 | 15.35±0.5 | 2.63±0.57 |

*Table S5* Amplitude Coefficients from fitting the HS1 resonance profiles using fixed energies and widths determined from fitting the WSe$_2$ $A_1'$ (249 cm$^{-1}$) profile. All amplitudes are scaled by a factor of $10^{14}$. Errors given for coefficients are a standard deviation obtained from the fitting process.

# Raman Peak Frequencies

For reference we provide the Raman shifts, obtained from fitting, for all peaks observed in monolayer WSe$_2$, and HS2 when resonant with the WSe$_2$ B1s exciton and for HS1 when resonant with the hybridised WSe$_2$ B1s and IX$^*$ excitonic states.

| | Raman Shift (cm$^{-1}$) | | |
|---|---|---|---|
| | **WSe$_2$** | **HS1** | **HS2** |
| | - | 38.2 ± 0.2 | - |
| | - | - | 106.3 ± 0.1 |
| | - | - | 123.9 ± 0.1 |
| | **218.3 ± 0.2** | 218.2 ± 0.5 | 218.0 ± 0.1 |
| | **222.0 ± 0.2** | 223.4 ± 0.3 | 222.0 ± 0.2 |
| | **234.8 ± 0.6** | 235.6 ± 0.6 | 234.6 ± 0.7 |
| | **240.6 ± 0.5** | 240.5 ± 0.2 | 239.9 ± 0.4 |
| | **249.3 ± 0.3** | 249.1 ± 0.1 | 249.2 ± 0.1 |
| | **257.8 ± 0.2** | 256.6 ± 0.4 | 256.1 ± 0.1 |
| | **261.6 ± 0.3** | 260.8 ± 0.4 | 260.6 ± 0.4 |
| | **283.5 ± 0.5** | - | - |
| | - | 291.0 ± 0.2 | - |
| | - | 309.2 ± 0.3 | - |
| | - | 353.7 ± 0.3 | 346.6 ± 0.7 |
| | **374.7 ± 0.3** | 374.3 ± 0.6 | 374.4 ± 0.2 |
| | **391.6 ± 0.2** | 390.9 ± 0.1 | 392.4 ± 0.1 |
| | **398.1 ± 0.2** | 396.4 ± 0.6 | 396.9 ± 0.2 |
| | **509.8 ± 0.5** | - | - |
| | **520 ± 0.1** | 520 ± 0.1 | 520 ± 0.1 |
| | **535.2 ± 0.2** | - | 532.5 ± 0.3 |

*Table S6 The fitted peaks positions are given for the different peaks observed in monolayer WSe$_2$, HS1 and HS2. These were determined by fitting multiple Raman spectra to a summation of Lorentzian line shapes. The errors given are a standard deviation obtained from fitting. All spectra were calibrated to the silicon peak at 520 cm$^{-1}$ which is shaded in grey.*

# Theoretical Modelling

## Estimation of the twist angle at which WSe₂ B1s and IX* become resonant.

The energy of the interlayer exciton that hybridises resonantly with the WSe₂ B1s, henceforth labelled IX*, has its minimum value $E^0_{IX^*}$ when its electron and hole are exactly at the WSe₂ and MoSe₂ valleys, respectively. That exciton has a finite momentum $\Delta K$, corresponding to the valley mismatch caused by the twist angle and the slightly different lattice constants of the two crystals. As such, this state is momentum dark. We are interested in the bright, zero-momentum IX* state of energy

$$E_{IX^*} = E^0_{IX^*} + \frac{\hbar^2 \Delta K^2}{2M_{IX^*}},$$

where $M_{IX^*}$ is the exciton mass, given by the sum of the WSe₂ conduction (CB) and MoSe₂ valence band (VB) masses (Table S7).

We begin by estimating the energy

$$E^0_{IX^*} = \delta_v + E^{WSe_2}_g - E^{IX^*}_b,$$

which requires knowledge of the offset between the VB edges of the two materials $\delta_v$; the WSe₂ electronic gap $E^{WSe_2}_g$; and the IX* exciton binding energy $E^{IX^*}_b$. We take $\delta_v$ from the *ab initio* literature (Table S8), and estimate $E^{IX^*}_b$ by numerically solving the two-body problem of a WSe₂ electron interacting with a MoSe₂ hole via the Keldysh potential [S7]

$$U(r) = -\frac{\pi e^2}{2\epsilon_{hBN} r_{eff}} \left[ H_0\left(\frac{r}{r_{eff}}\right) - Y_0\left(\frac{r}{r_{eff}}\right) \right].$$

Here, $e$ is the elementary charge, $\epsilon_{hBN} = 4$ is the average dielectric constant of the hBN encapsulation, and $r_{eff} = r_{WSe_2} + r_{MoSe_2} + d$ is the effective screening length for interlayer interactions [S8]. $r_{WSe_2} = 45.11$ Å and $r_{MoSe_2} = 39.79$ Å are the screening lengths of each individual layer [S9-10], and $d = 7.02$ Å is the interlayer separation, approximated by averaging the interlayer distances in bulk WSe₂ and MoSe₂ [S11-12]. The resulting value is reported in Table S8. Similarly, we solved the two-body problem for an electron-hole pair in the WSe₂ layer (setting $d = 0$ in $r_{eff}$, as in this case the electron and hole move in the same plane) and obtained the A1s exciton binding energy (Table S8). The WSe₂ band gap $E^{WSe_2}_g$ is then estimated by adding this value to the measured WSe₂ A1s exciton energy (see Supplementary Table S1). The final result is $E^0_{IX^*} = (2.039 \pm 0.111)$ eV.

Now we discuss the twist-angle dependence of the IX* kinetic energy. For small twist angles, the valley momentum mismatch is given by $\Delta K = 4\pi\theta(1-\delta)/(3a_{WSe_2})$, where $\theta$ is the twist angle in radians, $a_{WSe_2}$ is the corresponding lattice constant, and the lattice mismatch

$$\delta = 1 - \frac{a_{WSe_2}}{a_{MoSe_2}} = (3.040 \pm 0.006) \times 10^{-3},$$

is negligible for this chalcogen-matched heterostructure. The kinetic energy is then given by

$$\frac{\hbar^2 \Delta K^2}{2M_{IX^*}} = \frac{\hbar^2}{2M_{IX^*}} \left(\frac{4\pi}{3a_{WSe_2}}\right)^2 \theta^2.$$

Resonance between the momentum bright IX$^*$ and WSe$_2$ B1s excitons is achieved for a twist angle

$$\theta = \frac{3a_{WSe_2}}{4\pi}\sqrt{\frac{2M_{IX^*}}{\hbar^2}\Delta E},$$

determined by the energy offset between the two excitons, $\Delta E = E_{WSe_2\,B1s} - E^0_{IX^*} = (125 \pm 112)$ meV (see Supplementary Table S2). Measured from 60° alignment, the resulting twist angle is $51.1° \pm 8.9°$. For reference, the above analysis gives a $(108 \pm 116)$ meV detuning between the two exciton energies for sample HS1, with a relative angle of 57°. This is consistent with the $(46.9 \pm 0.43)$ meV value inferred from reflectance measurements; see Fig. 5 in the main text.

**Tunnelling strength estimation from the reflectance contrast measurements.**

To model the hybridisation of WSe$_2$ B1s and IX$^*$ by interlayer hole tunnelling, we used the exciton wavefunctions obtained in the numerical solution described above. The exciton Bohr radii $a_{B1s}$ and $a_{IX'}$ were extracted fitting the numerical solutions with *1s*-type trial wave functions [10]. Then, following Ref [S13], the exciton mixing strength is given by

$$T_{B1s,IX^*} = -\frac{4t_v}{a_{B1s}a_{IX^*}}\left(\frac{a_{B1s}+a_{IX^*}}{a_{B1s}a_{IX^*}}\right)\left[\left(\frac{a_{B1s}+a_{IX^*}}{a_{B1s}a_{IX^*}}\right)^2 + \left(\frac{m^e_{WSe_2}}{m^e_{WSe_2}+m^h_{MoSe_2}}\right)^2 \Delta K^2\right]^{-3/2},$$

where $t_v$ is the interlayer valence-band tunnelling energy, and $m^e_{WSe_2}$ ($m^h_{MoSe_2}$) is the WSe$_2$ electron (MoSe$_2$ hole) mass (see Table S7). The hybridised exciton energies are two of the eigenvalues of the intralayer-interlayer exciton Hamiltonian with unknown exciton energies $\varepsilon_{B1s}$ and $\varepsilon_{IX^*}$, and hopping energy $T_{B1s,IX^*}$, given by

$$\varepsilon_\pm = \left(\frac{\varepsilon_{B1s}+\varepsilon_{IX^*}}{2}\right) \pm \sqrt{3T^2_{B1s,IX^*} + \left(\frac{\varepsilon_{B1s}-\varepsilon_{IX^*}}{2}\right)^2}.$$

Fixing $\varepsilon_\pm$ to the values inferred from the resonance Raman data analysis, this equation can be solved for $\varepsilon_{B1s}$ and $\varepsilon_{IX^*}$ as a function of $t_v$. This is shown by the solid curves in Fig. 5(b).

To evaluate the relative (integrated) absorption rates of the two hybridised excitons, we assume a vanishing oscillator strength for IX$^*$, such that optical activity of either hybridised exciton must come from its corresponding WSe$_2$ B1s component, computed during the above solution of the hybridisation problem. Then, following Ref [S13], we obtain

$$\frac{I_+}{I_-} = \frac{\varepsilon_+ - \frac{\varepsilon_{B1s}-\varepsilon_{IX^*}}{2}}{\varepsilon_+ + \frac{\varepsilon_{B1s}-\varepsilon_{IX^*}}{2}}.$$

This ratio is plotted in Fig. 5(c) as a function of $t_v$. The calculated absorption spectra shown in Fig. 5(b) follow this ratio and were generated using the line energies and broadening factors extracted from the resonance Raman data, assuming Lorentzian line shapes.

Conversely, we may fix the WSe$_2$ B1s energy, 57° twist angle, and the tunnelling energy $t_v = 11.4$ meV determined in the main text, and vary the IX$^*$ energy as one may accomplish by means of an out-of-plane electric field. This is explored by the absorption spectrum of Fig. S8. The excitonic spectrum was computed following Ref. [S13], using a plane wave basis folded into the mini-Brillouin zone defined by the moiré superlattice. The signature of strong moiré effects is the presence of a fine structure in the optical spectrum, emerging from the exciton moiré mini bands. Instead, the spectrum of Fig. S8 shows a simple avoided crossing as a function of the IX* Stark shift, indicating that moiré effects in the hybridised exciton formation are weak in this energy range.

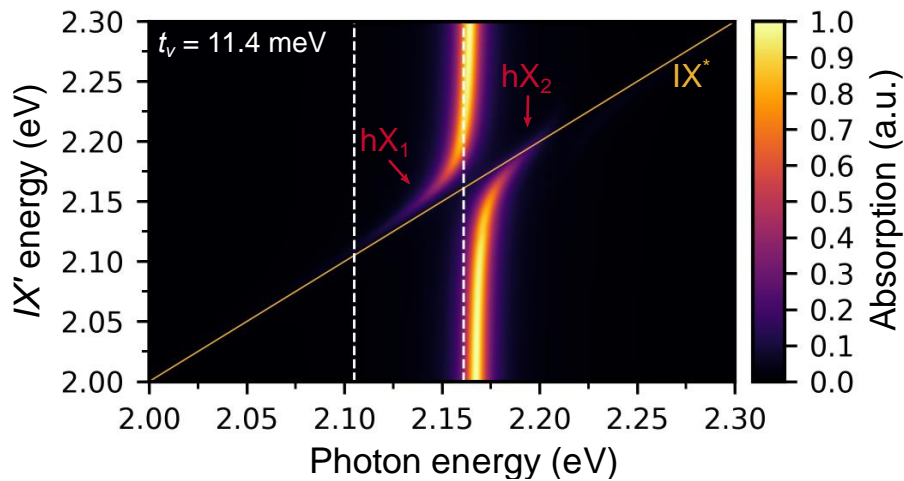

**Figure S8** *Calculated absorption spectrum of 57° MoSe$_2$/WSe$_2$ around the WSe$_2$ B1s line as a function of the bare IX* energy, for tunnelling strengths 11.4meV. Close to resonance, the two excitons mix to form the hybridized excitons hX$_1$ and hX$_2$.*

| | Lattice constant (Å) | CB SO splitting (meV) | VB SO splitting (meV) | CB mass ($m_0$) | VB mass ($m_0$) | Screening length (Å) |
|---|---|---|---|---|---|---|
| WSe$_2$ | 3.284±0.002[a,b] | -37.0±0.01[c] | 485.0 ± 10.0[d] | 0.50 ± 0.10[e] | 0.42±0.05[d] | 45.11[f,g] |
| MoSe$_2$ | 3.294±0.005[a,b] | 21.0 ± 0.10[c] | 220.0 ± 30.0[d] | 0.80 ± 0.20[e] | 0.50±0.10[d] | 39.79[f,g] |

**Table S7** *Lattice and band structure parameters of monolayer WSe$_2$ and MoSe$_2$. All values have been extracted from the experimental literature, except for the conduction-band spin-orbit splittings at the K point and the screening lengths, which are ab initio results. Valence band masses and spin-orbit splittings are based on ARPES experiments; the conduction-band masses were obtained from magnetotransport data. References: a) Wilson and Yoffe (1969) [S14]; b) Al-Hilli and Evans (1972) [S11]; c) Kormányos et al. (2015) [S15]; d) Nguyen et al. (2019) [S16]; e) Gustafsson et al. (2018) [S17]; f) Kumar and Ahluwalia (1972) [S9]; g) Berkelbach et al. (2013) [S10].*

| | Interlayer distance (Å) | Screening length $r_{eff}$ (Å) | VB detuning $\delta_v$ (eV) | WSe$_2$ A1s binding energy (meV) | IX* binding energy (meV) |
|---|---|---|---|---|---|
| MoSe$_2$/WSe$_2$ | 7.02[a,b] | 91.39 | 0.306 ± 0.110[c,d,e] | 136 | 140 |

**Table S8** *Electronic and excitonic band structure parameters of the MoSe$_2$/WSe$_2$ heterostructure. For the interlayer distance, we averaged those measured experimentally for bulk MoSe$_2$ and WSe$_2$, whereas the valence-band edge offset $\delta_v$ was obtained by comparing three different density functional theory approximations available in the literature. The exciton binding energies were computed by solving the corresponding two-body problems, described in the text. References: a) Al-Hilli and Evans (1972) [S11]; b) Hicks (1964) [S12]; c) Kang et al. (2013) [S18]; d) Gong et al. (2013) [S19]; e) Xu et al. (2018) [S20].*